\documentclass[12pt]{iopart}

\usepackage[english]{babel}
\usepackage[dvips]{graphicx}
\usepackage{amsfonts}
\usepackage{amssymb,latexsym}
\usepackage{mathrsfs}
\usepackage{epsfig}
\usepackage{subfigure}

\newcommand{\be}{\begin{eqnarray}}
\newcommand{\ee}{\end{eqnarray}}
\newcommand{\bdm}{\begin{displaymath}}
\newcommand{\edm}{\end{displaymath}}
\newcommand{\ba}{\begin{array}}
\newcommand{\ea}{\end{array}}

\providecommand{\ti}{{\mathrm{t}}}
\def\Orb{{\mathbf{S}^1/\mathbf{Z}_2}}
\def\Z2{{\mathbf{Z}_2}}
\def\mX{{\mathbf{X}}}

\begin{document}

\title{Large Gravitational Waves and Lyth Bound in Multi Brane Inflation}
\author{Axel Krause}
\address{Arnold Sommerfeld Center for Theoretical Physics,
Department f\"ur Physik, Ludwig-Maximilians-Universit\"at M\"unchen,
Theresienstr.~37, 80333 M\"unchen, Germany}
\eads{\mailto{axel.krause@physik.uni-muenchen.de}}

\begin{abstract}
It is shown that multi M5-brane inflation in heterotic M-theory gives rise to a detectable gravitational wave power spectrum with tensor fraction $r$ typically larger than the projected experimental sensitivity, $r_{exp} = 0.01$. A measurable gravitational wave power spectrum entails a large inflationary energy scale and a super-Planckian inflaton variation. They present serious problems for particle theory model building resp.~a reliable effective field theory description. These problems are eased or even absent in multi-brane inflation models and multi M5-brane inflation, in particular.
\end{abstract}

\noindent{\it Keywords\/}: Gravitational Waves, Inflation, Heterotic M-Theory\\
{\it ArXiv ePrint\/}: 0708.4414\\
{\it Preprint id\/}: \verb+JCAP_051P_0108+

\maketitle
\tableofcontents

\section{Introduction}

String-theory remains, to date, without direct experimental support. It encompasses a host of concepts, ranging from supersymmetry to extra dimensions. A discovery of one of which at the LHC would certainly point strongly towards string-theory but, yet, would not provide conclusive evidence. Unless the fundamental string-scale happens to be extremely low, not far above the TeV scale, it seems impossible to test string-theory directly at colliders. Cosmological observations, on the other hand, are sensitive to physics at very high energy scales, such as the Grand Unification (GUT) scale. For instance, quantum fluctuations of the inflaton during inflation, lead to observable signatures in the power spectrum of density perturbations, revealing clues about the underlying physics at those early epochs.
Great effort has therefore been invested over the past few years to connect string-theory to cosmic inflation and to determine its cosmological signatures (see \cite{Quevedo:2002xw} for reviews). For instance, the end of brane-inflation is characterized by the production of cosmic superstrings \cite{Sarangi:2002yt}, \cite{Jones:2003da}, \cite{Copeland:2003bj}, \cite{Becker:2005pv} whose interactions could lead to observational signatures different from those of conventional cosmic strings \cite{Jackson:2004zg}, \cite{Jackson:2006rb}, \cite{Jackson:2007hn}. Whereas the former are fundamental objects, the latter merely represent effective descriptions of field theory vortex solutions. The production of cosmic superstrings could also serve as a tool to differentiate observationally between various string-inflation scenarios \cite{Pogosian:2003mz}, \cite{Shlaer:2005ry}, \cite{Leblond:2007tf}.

Another distinctive signature could be gravitational waves, which are generated during inflation and might be observable in the low multipoles of the Cosmic Microwave Background (CMB) anisotropy. Their detection would directly reveal the value $V$ of the inflaton potential together with its slope $V'$ \cite{Liddle:1993fq}. In association with measurements of the spectral index of density perturbations it would also constrain the curvature $V''$ \cite{Liddle:1992wi}. Gravitational waves do not couple to density fluctuations of the universe, and thus are not responsible for its large-scale structure. But they induce fluctuations in the CMB which provide a unique signature of inflation.

Related to a possible future detection of gravitational waves from inflation are two immediate enigmas. The first is the implied large value of $V$ around the GUT scale, the second is the Lyth bound \cite{Lyth:1996im}. The Lyth bound states that the inflaton field variation has to exceed the reduced Planck scale $M_{Pl}$ $=\sqrt{8\pi G} = 2.4\times 10^{18}$GeV if gravitational waves become detectable. This would render an effective field theory description unreliable if couplings take natural values of order one.

For a variety of string inflation models it has been demonstrated that gravitational wave amplitudes are too small \cite{Baumann:2006cd}, \cite{Bean:2007hc}, \cite{Kallosh:2007wm}, \cite{Lidsey:2007gq}, \cite{Peiris:2007gz} to become observable by planned CMB polarization experiments \cite{Taylor:2004hha}, \cite{Verde:2005ff}. A potential future detection of tensor modes from inflation would therefore rule all these string inflation models out in one stroke. Do we really face a No-Go, stating that all string inflation scenarios cannot generate detectable gravitational waves? Our goal in this paper is twofold. First, we answer this important question in the negative, by demonstrating that the multi M5-brane inflation scenario of \cite{Becker:2005sg} gives rise to sizeable gravitational waves, which ought to be easily detectable by the Planck satellite. This confutes the No-Go. Second, we address the two implied problems, mentioned in the previous paragraph, which arise when gravitational waves from inflation become detectable. We show that the multi M5-brane inflation scenario of \cite{Becker:2005sg} leads to  parametric changes in both constraints which allow the two problems to be mitigated.

\section{Scalar and Tensor Perturbations}

For the definition of the relevant cosmological parameters it suffices to adopt an effective single inflaton model, specified by an inflaton potential $V(\phi)$, resp.~Hubble parameter $H(\phi)$, with canonically normalized inflaton $\phi$. The amplitudes $A_S(k)$ and $A_T(k)$ of the scalar and tensor power spectra generated by inflation are defined to lowest order in slow-roll parameters according to the conventions of \cite{Stewart:1993bc}, \cite{Lidsey:1995np} by\footnote{$A_S$ coincides with $\delta_H$ as defined in \cite{Liddle:1993fq}.}
\be
A_S(k) = \frac{1}{10\pi} \frac{H^2}{M_{Pl}^2|H'|}\bigg|_{k=aH} \; , \qquad
A_T(k) = \frac{1}{5\sqrt{2}\pi} \frac{H}{M_{Pl}}\bigg|_{k=aH} \; .
\ee
The scalar and tensor spectral indices, evaluated at a ``pivot'' scale $k_0 = (500{\rm Mpc})^{-1}$ \cite{Peiris:2003ff}, follow as
\be
n_S - 1 = \frac{d\ln A_S^2(k)}{d\ln k}\bigg|_{k=k_0} \; , \qquad
n_T = \frac{d\ln A_T^2(k)}{d\ln k}\bigg|_{k=k_0} \; .
\ee
The important tensor-to-scalar ratio $r$ is given, in the conventions of \cite{Peiris:2003ff}, by
\be
r = 16 \left(\frac{A_T}{A_S}\right)^2 \; .
\ee
In slow-roll inflation the last three quantities possess simple expressions
\be
n_S - 1 = -6\epsilon+2\eta \\
n_T = -2\epsilon \\
r = 16\epsilon
\ee
to lowest order in the slow-roll parameters\footnote{Alternative parameters which are often used are,
$\epsilon_H = 2 M_{Pl}^2 (H'/H)^2$, $\eta_H = 2M_{Pl}^2 H''/H$. They are related through $\epsilon_H = \epsilon$, $\eta_H = \eta - \epsilon$ in slow-roll approximation.} $\epsilon = (M_{Pl}^2/2) \left(V'/V\right)^2$ and $\eta = M_{Pl}^2 V''/V$. The ensuing relation, $n_T = -r/8$, is known as the lowest order consistency relation. It is thus three independent parameters, $A_S, A_T, n_S$, which fully characterize an inflation model to lowest order. A potential running of the scalar spectral index, $dn_S/d\ln k \ne 0$, is based on higher-order effects.

The gravitational wave power spectrum, ${\cal P}_T = 100 A_T^2$, \cite{Stewart:1993bc}, \cite{Lidsey:1995np} is thus suppressed by the small slow-roll parameter $\epsilon$ compared to the scalar power spectrum, ${\cal P}_S = (25/4) A_S^2$ \cite{Stewart:1993bc}, \cite{Lidsey:1995np}. In most string-inflation scenarios one finds a hierarchy, $\epsilon \ll \eta \ll 1$, implying a very small $r$ and thus undetectable gravitational waves. The multi M5-brane inflation scenario of \cite{Becker:2005sg}, on the contrary, leads to $\epsilon \approx \eta \ll 1$, hence indicating a larger $r$. The present best experimental upper limit on $r$ depends on whether running of $n_S$ is allowed or excluded. Combining Wilkinson Microwave Anisotropy Probe 3-year (WMAP3) and Sloan Digital Sky Survey (SDSS) data one finds $r<0.31$ and $0.93<n_S<1.01$ in the absence of running and $r<0.38$ together with $0.97<n_S<1.21$ and $-0.13< dn_S/d\ln k <0.007$ in the presence of running \cite{Kinney:2006qm} (see also \cite{Chongchitnan:2006pe}, \cite{Efstathiou:2006ak}, \cite{Friedman:2006zt}, \cite{Xia:2007gz}).

\section{Multi M5-Brane Inflation}

Since we are interested in the production of gravitational waves from multi M5-brane inflation in heterotic M-theory, let us briefly review its main features (for details the reader is referred to \cite{Becker:2005sg}).

The ${\cal N}=1$ supersymmetric compactification of heterotic M-theory down to four dimensions takes place on a seven-manifold $\mX\times\Orb$, built out of a Calabi-Yau threefold $\mX$ which is fibered, via a warp-factor, along the orbifold interval $\Orb$ \cite{Witten:1996mz}, \cite{Curio:2000dw}. Cancelation of tadpoles in heterotic M-theory requires generically to introduce $N$ additional M5-branes parallel to the $\Orb$ orbifold boundaries. Classically, there is no interaction among the M5-branes, but quantum-mechanically the M5-branes interact via non-perturbative open M2-instanton exchange \cite{Moore:2000fs}, \cite{Lima:2001jc}, \cite{Curio:2001qi}. The leading order interaction stems from $N-1$ open M2-instantons, each wrapping a holomorphic 2-cycle on $\mX$ and stretching a distance $\Delta x_n = x^{11}_{n+1}-x^{11}_n > 0\,$; $n=1,\ldots,N-1$ along $\Orb$, such as to connect neighboring M5-branes at positions $x_n^{11}$ and $x_{n+1}^{11}$. The geometry requires $0 \le x_n^{11} \le L$, with $L$ being the $\Orb$ length. Supersymmetry is broken spontaneously by the M2-instantons. A priori, the setup gives rise to a multi-inflaton model, after identifying the $N-1$ distance moduli, $\Delta x_n$, with inflatons. Since the identification, $\Delta x \equiv \Delta x_1 = \ldots = \Delta x_{N-1}$, turns out to correspond to a stable attractor solution in this model, one ends up with an effective single inflaton model in which $\Delta x$ represents the inflaton. It is related to the canonically normalized inflaton, $\varphi$, via \cite{Becker:2005sg}
\be
\frac{\varphi}{M_{Pl}} = t \sqrt{\frac{p_N}{2}} \frac{\Delta x}{L}
\; .
\label{Variation}
\ee
The orbifold-length modulus, $t$, relates to the dilaton in the weakly coupled limit ($L\rightarrow 0$), $s$ denotes the Calabi-Yau volume modulus, and
\be
p_N = N(N^2-1)\left(\frac{4}{3st}\right)
\ee
is a moduli dependent parameter which controls the dynamics of the effective potential \cite{Becker:2005sg}. It scales like $p_N\sim N^3$ at large $N$.

The stabilization of $s$ arises from an interplay in the hidden sector of the gaugino condensate with M5-instantons \cite{Curio:2006dc} (a similar stabilization of K\"ahler moduli applies to the heterotic string \cite{Curio:2005ew}), whereas the $t$ modulus is fixed either by balancing in the hidden sector the $H$-flux with the gaugino condensate \cite{Krause:2007gj} or by an interplay of the gaugino condensate with open M2-instantons stretching from boundary to boundary \cite{Becker:2004gw} (see also \cite{Buchbinder:2003pi}, \cite{Buchbinder:2004im}, \cite{Anguelova:2005jr}, \cite{Anguelova:2006qf}, \cite{Braun:2006th}, \cite{Correia:2006vf}, \cite{Gray:2007mg}, \cite{Correia:2007sv}). In both cases the generated vacuum energy is small compared to the one driving inflation and is therefore expected not to alter the inflationary dynamics. Subsequently, we are treating both moduli as fixed parameters and choose values $s = 800$ and $t = 2$ for exemplification. A full discussion of the stabilized values in this context is left to a separate publication. The effective inflaton potential which emerges in the large volume limit, specified by $st > y^2$ with $y^2 = \sum_{n=1}^N y_n^2$ and $y_n = t x_n^{11}/L$, is \cite{Becker:2005sg}
\be
V(\varphi) = (N-1)^2 V_{M5}(\varphi) \; ,
\ee
where
\be
V_{M5}(\varphi) = V_0 e^{-\sqrt{\frac{2}{p_N}} \frac{(\varphi-\varphi_i)}{M_{Pl}}} \; .
\ee
The prefactor is\footnote{In our conventions $i\int_{\mX} \Omega \wedge \overline{\Omega} = 1$.} $V_0 = 6M_{Pl}^4/(s t^3 d)$, $d$ being the intersection number of $\mX$ for which we assume a Hodge number $h^{(1,1)}(\mX)=1$. The value of the inflaton at initial time $\ti_i$ is denoted by $\varphi_i = \varphi(\ti_i)$. Note that the inflationary potential is enhanced by an M5-brane dependent factor, resulting in a scaling, $V(\varphi) \sim N^2$, at large $N$.
The ensuing Friedmann-Robertson-Walker (FRW) cosmological evolution induced by this potential is known as power-law inflation \cite{Lucchin:1984yf}. It exhibits a scale-factor
\be
a(\ti) = a_i \left(\frac{\ti}{\ti_i}\right)^{p_N}
\ee
and inflaton evolution
\be
\varphi(\ti) = \varphi_i+\sqrt{2p_N} M_{Pl}
\ln \left(\frac{\ti}{\ti_i}\right) \; ,
\ee
with initial time given by
\be
\ti_i = M_{Pl} \sqrt{\frac{(3p_N-1)p_N}{(N-1)^2 V_0}} \; .
\ee

The reason why the steep exponential potentials, generated by the open M2-instanton exchanges, give rise to inflation is that sufficiently many of them generate a large Hubble friction which drives the system into the slow-roll regime, a mechanism known as assisted inflation \cite{Liddle:1998jc}. This phenomenon is clearly visible in the slow-roll parameters
\be
\epsilon = \frac{1}{p_N} \sim \frac{1}{N^3} \; , \qquad
\eta = \frac{2}{p_N} \sim \frac{1}{N^3} \; ,
\ee
which become small when $N$ turns large. Although power-law inflation is not viable per se since it continues forever (as evident from $\epsilon$ and $\eta$ being constant), this is not true for multi M5-brane inflation. Here, towards the end of inflation, the M5-branes collide successively with the $\Orb$ boundaries. This process decreases $N$ in discrete steps, thus causes $p_N$ to decrease and finally terminates inflation as soon as $p_N$ drops below 1 \cite{Ashoorioon:2006wc}.

\section{Detectable Gravitational Waves}

The above slow-roll parameters imply a red spectrum for multi M5-brane inflation
\be
n_S = 1 - \frac{2}{p_N} \; ,
\ee
and for the tensor modes
\be
n_T = -\frac{2}{p_N} \; , \qquad r = \frac{16}{p_N} \; ,
\ee
at leading linear order in slow-roll. A non-trivial running of $n_S$ shows up only at quadratic order, which requires the third slow-roll parameter
\be
\xi^2 \equiv M_{Pl}^4 \frac{V' V'''}{V^2} = \frac{2}{p_N}
\ee
to be included \cite{Kosowsky:1995aa}. The result is no running
\be
\frac{d n_S}{d \ln k} = -16\epsilon\eta+24 \epsilon^2+2\xi^2 = 0
\ee
since $p_N$ is constant. Towards the end of inflation $p_N$ decreases \cite{Ashoorioon:2006wc} and thus induces a running of $n_S$. This is, however, without consequences for the detection of gravitational waves because the scales on which the tensor modes are observed leave the horizon already after 4 e-foldings of inflation \cite{Lyth:1996im}, \cite{Alabidi:2006fu}. The observational bounds from the combined WMAP3 and SDSS data, which apply in the absence of running, are $r < 0.31$ and $0.93 < n_S < 1.01$ (WMAP3 alone gives the weaker bound, $r < 0.6$, for about the same $n_S$ range) \cite{Kinney:2006qm}. Moreover, $r$ is clearly correlated with $n_S$ \cite{Spergel:2006hy}, see fig.\ref{GravWave} below. Thus, a low $n_S = 0.95$ is valid for $r < 0.05$, whereas a larger, e.g.~$n_S = 0.98$, holds for $r \approx 0.15$.

From the above formulae we find that in multi M5-brane inflation
\be
r = \frac{12 st}{N(N^2-1)} \approx \left(\frac{26.8}{N}\right)^3 \; ,
\ee
where in the last relation the aforementioned values for $s$ and $t$ were used. The number of M5-branes is determined, by adopting a reasonable value for $n_S$. For instance, choosing $n_S = 0.98$, because a detectable $r$ is generated in multi M5-brane inflation, as we will soon see,
fixes $p_N = 100$. The latter has to equal
\be
p_N \approx \left(\frac{N}{10.6}\right)^3 \; ,
\ee
which results from a substitution of the specified values for $s$ and $t$. Hence, in this example, of order $N \approx 49$ M5-branes are needed to account for $n_S = 0.98$ and the ensuing substantial tensor fraction of $r = 1.6\times 10^{-1}$.

It is interesting to relate the scalar spectral index and tensor fraction through elimination of the model-dependent parameter $p_N$. This yields the simple linear relation
\be
r = 8(1-n_S) \; ,
\ee
which is plotted and compared to WMAP3 constraints in fig.\ref{GravWave}. First, the plot demonstrates that $r$ in multi M5-brane inflation is generically larger than the projected experimental sensitivity of $r=10^{-2}$ and should therefore allow a detection by the Planck satellite. In contrast to the ordinary power-law inflation $r$-$n_S$ relation \cite{Smith:2005mm}, multi M5-brane inflation places a lower bound on $r$, which arises as follows. The condition, $st > y^2$, under which the derivation of the exponential inflaton potential from the M5-brane dynamics is valid, yields an upper bound on $N$. This translates into an upper bound on $p_N$ resp.~a lower bound on $r$. For instance, for a maximal $N$ of order 200 \cite{Becker:2005sg} one finds, with values for $s$ and $t$ as before, a lower bound $r \ge 2.4\times 10^{-3}$, which is close to the experimental sensitivity. This refutes the No-Go, stated above, that string-theory models for inflation cannot generate detectable gravitational waves. Second, the plot in fig.\ref{GravWave} demonstrates that the straight $r$-$n_S$ line passes right through the center of the parameter region allowed by WMAP3 observation. The constraint, $r < 0.31$, from WMAP3 plus SDSS puts a lower bound on the number of M5-branes present during inflation, of order $N > 40$. Let us emphasize that, even though $n_S$ and $r$ depend each, via $p_N$, on M-theory parameters, such as $s,t,N$, the above linear relation is free of any such parameters and constitutes a firm falsifiable prediction of the model.

Interestingly, a detection of a non-zero $r$ by future CMB experiments would immediately place an upper bound on the logarithmically averaged effective equation-of-state parameter, $\hat w$, from the end of inflation until the start of big bang nucleosynthesis, as has been shown in \cite{Boyle:2007zx}.
\begin{figure}[tb]
\includegraphics[scale=0.7]{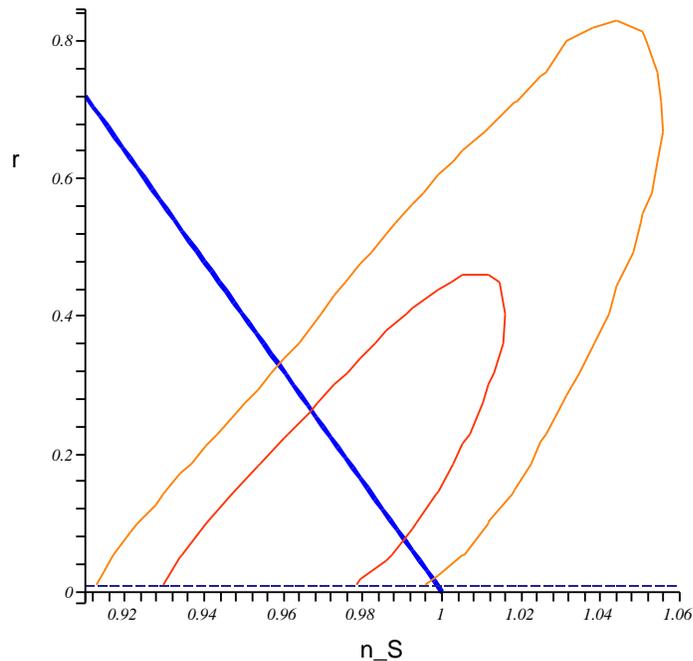}
\caption{\label{GravWave} The straight inclined line displays the $r$-$n_S$ relation resulting from multi M5-brane inflation, which terminates at some finite lower value $r_{low} > 0$. It shows that $r$ is in general larger than the anticipated observational threshold $r_{exp} = 0.01$ (horizontal dashed line) and that the line goes right through the allowed central region. The parameter region inside the big (small) ellipse is allowed by WMAP3 data at 95\% (66\%) confidence level.}
\end{figure}

\section{Lyth Bound in Multi-Brane Inflation}

With detectable gravitational waves generated by inflation, the low-energy effective theory description is known to exhibit two generic problems -- the large inflationary energy-scale and the Lyth bound, implying super-Planckian variation of the inflaton. Both problems are mitigated when the microscopic multi-brane origin of the effective theory is taken into account, as we will discuss next.

The amplitude of the gravitational wave CMB anisotropy fixes the energy-scale of slow-roll inflation at \cite{Lyth:1984yz}
\be
V^{1/4} = 3.3 \times 10^{16} r^{1/4} \, {\rm GeV} \; .
\ee
Hence, detectable gravitational waves with $r > 0.01$ yield a large inflation energy-scale of $V^{1/4} > 10^{16}\,$GeV, which is difficult to obtain from particle theory models. They would favor smaller energy-scales e.g.~$(M_{EW} M_{Pl})^{1/2}$ or $(M_{EW} M_{Pl})^{1/3}$, where $M_{EW} = 100\,$GeV \cite{Lyth:1996im}. By going beyond the effective power-law inflation description and employing the M-theory origin of multi M5-brane inflation, this problem is easily alleviated. Namely, one notices that $V = (N-1)^2 V_{M5}$ is enhanced by the non-dynamical M5-brane factor $(N-1)^2$ and the ``true'' energy-scale $V_{M5}$ is thus fixed at a parametrically lower value
\be
V_{M5}^{1/4} = \frac{3.3 \times 10^{16} r^{1/4}}{(N-1)^{1/2}} \, {\rm GeV} \; .
\ee
Of course, there is no problem in our case with ordinary particle theory model building as the results were consistently derived from M-theory \cite{Becker:2005sg}. Similar conclusions hold for all multi-brane inflation models \cite{ToAppear} or the string-theory realization \cite{Olsson:2007he} of N-flation \cite{Dimopoulos:2005ac}.

The second problem arises from the Lyth bound \cite{Lyth:1996im}. This bound arises in slow-roll inflation where the variation of the inflaton, measured in terms of e-foldings ${\cal N}_e$,
\be
\frac{d\varphi}{d{\cal N}_e} = M_{Pl} \sqrt{2\epsilon}
\ee
is combined with $r=16\epsilon$ to give, after integration over the full inflationary period,
\be
\frac{\Delta\varphi}{M_{Pl}} = \frac{1}{\sqrt{8}} \int_0^{{\cal N}_e} d{\cal N}_e \sqrt{r}  >  \frac{1}{\sqrt{8}} \int_0^{{\cal N}_{CMB}} d{\cal N}_e \sqrt{r}  \ge  \sqrt{2r} \; .
\ee
This is a rather conservative bound\footnote{A slightly tighter bound is derived in \cite{Boubekeur:2005zm} which finds $\Delta\varphi/M_{Pl} > {\cal N}_e \sqrt{r/8}$. In fact, for our case of constant slow-roll parameter $\epsilon$ and hence constant $r$, one can perform the integral exactly with the result,
$\Delta\varphi/M_{Pl} = {\cal N}_e \sqrt{r/8}$, which saturates the latter bound. An even stronger bound, which has been derived in \cite{Boubekeur:2005zm} for quadratic hilltop potentials, is not compatible with the last equality and hence does not apply in our case.}, which uses an estimate based on the fact that ${\cal N}_{CMB} = 4$ e-foldings pass until wavelengths corresponding to CMB multipoles $2\le l\le 100$ cross the Hubble radius. The Lyth bound implies that large field inflation models, which are characterized by a total super-Planckian inflaton variation, $\Delta \varphi \ge M_{Pl}$, are correlated with large tensor modes. However, their description in terms of an effective field theory
\be
V = V_{vac} + \frac{1}{2}m^2\varphi^2 + \lambda \varphi^4 + M_{Pl}^4 \sum_{i=3,4,\ldots} \lambda_{2i} \left(\frac{\varphi}{M_{Pl}}\right)^{2i} \; ,
\ee
becomes questionable, since higher-order corrections cease to be small\footnote{\cite{Easther:2006qu} shows that also small-field models may face the same problem.}. One possibility is to endow the non-renormalizable couplings $\lambda_{2i}$ with much smaller values than the natural ${\cal O}(1)$ values, as in chaotic inflation \cite{Linde:2004kg}.

In multi M5-brane inflation the inflaton $\varphi$ is only an effective field, and the fundamental microscopic field, which needs to be considered, is the modulus $\Delta x$. Using equation (\ref{Variation}), the Lyth bound, in terms of $\Delta x$, turns into
\be
\frac{\Delta x_f}{L} \ge \left(\frac{1}{\sqrt{p_N} t}\right) 2\sqrt{r} \; ,
\ee
where we have approximated the total variation $\Delta x_f - \Delta x_i$ during inflation by the final value $\Delta x_f$, since $\Delta x_f \gg \Delta x_i$, see \cite{Becker:2005sg}. Hence, the Lyth bound becomes parametrically weakened, since $\sqrt{p_N}\sim N^{3/2}$. Moreover, the problematic original bound, which implied large super-Planckian energy densities, turns into a bound which requires sufficiently large distances $\Delta x_f$. This is, however, perfectly compatible with an effective supergravity description which is valid in the long rather than the short wavelength approximation to M/string-theory. Finally, evaluating the bound, e.g.~for $s=800, t=2, N = 49$, leads to $\Delta x_f/L \ge \sqrt{r}/10$ which is well consistent with the geometric orbifold constraint, requiring $\Delta x_f \le L$.

In fact the latter parametric weakening of the Lyth bound plus large energy density to large distance transformation is characteristic of all multi-brane inflation models, even though they might not generate sizeable gravitational waves. In multi-brane inflation models \cite{ToAppear} the multi-inflaton description can often be reduced to an effective single-inflaton description, depending on the amount of symmetry of the multi brane setup. Given this is the case, the effective canonically normalized inflaton $\varphi$ is related to the microscopic open string distance modulus $\Delta X$, which describes e.g.~the common distance between adjacent branes in a hypercubic lattice arrangement, via \cite{ToAppear}
\be
\frac{\varphi}{M_{Pl}} = \left(\frac{c_N}{M_{Pl}}\right) \Delta X
\; .
\ee
The renormalization coefficient $c_N$, which depends on the number of interacting branes, $N$, scales like \cite{ToAppear}
\be
c_N \sim N^b \; , \quad b>0
\ee
with positive scaling exponent $b$, depending on the way in which the branes are distributed over the compactification manifold. E.g.~for a uniform distribution over a $d$-dimensional hypercubic lattice one finds, $b = 3/(2d)$ \cite{ToAppear}. The Lyth bound is then parametrically weakened to
\be
|\Delta X_f - \Delta X_i| \ge \left(\frac{M_{Pl}}{c_N}\right) \sqrt{2r} \sim \frac{1}{N^b} \sqrt{2r} \; ,
\ee
and furthermore transformed from a large energy to a large distance bound, in compliance with any long wavelength approximation to M/string-theory. This weakening of the Lyth bound should hold similarly for $N$-flation \cite{Kinney:2007np} and other assisted inflation models, such as~\cite{Singh}. In the special case, however, of warped D3-brane inflation in a Klebanov-Strassler throat, with all $N$ D3-branes coinciding and under neglect of the backreaction of the D3-branes, the effective inflaton range stays upper-bounded by $M_{Pl}$ \cite{Huang:2007qz}.

\ack
We thank Liam McAllister for discussions and Laila Alabidi for help with WMAP3 data. This work has been supported by the DFG and the Transregional Collaborative Research Centre TRR~33 ``The Dark Universe''.


\section*{References}
\begin{thebibliography}{999}

\bibitem{Quevedo:2002xw}
  F.~Quevedo,
  Class.\ Quant.\ Grav.\  {\bf 19}, 5721 (2002)
  [arXiv:hep-th/0210292].\\
  C.~P.~Burgess,
  Pramana {\bf 63}, 1269 (2004)
  [arXiv:hep-th/0408037].\\
  A.~Linde,
  eConf {\bf C040802}, L024 (2004)
  [J.\ Phys.\ Conf.\ Ser.\  {\bf 24}, 151 (2005\ PTPSA,163,295-322.2006)]
  [arXiv:hep-th/0503195].\\
  C.~P.~Burgess,
  arXiv:hep-th/0606020.\\
  S.~H.~Henry Tye,
  arXiv:hep-th/0610221.\\
  J.~M.~Cline,
  arXiv:hep-th/0612129.\\
  R.~Kallosh,
  arXiv:hep-th/0702059.\\
  M.~Gasperini and G.~Veneziano,
  arXiv:hep-th/0703055.\\
  A.~Linde,
  arXiv:0705.0164 [hep-th].\\
  J.~M.~Cline,
  arXiv:0705.2982 [hep-th].\\
  C.~P.~Burgess,
  arXiv:0708.2865 [hep-th].

\bibitem{Sarangi:2002yt}
  S.~Sarangi and S.~H.~H.~Tye,
  Phys.\ Lett.\  B {\bf 536}, 185 (2002)
  [arXiv:hep-th/0204074].

\bibitem{Jones:2003da}
  N.~T.~Jones, H.~Stoica and S.~H.~H.~Tye,
  Phys.\ Lett.\  B {\bf 563}, 6 (2003)
  [arXiv:hep-th/0303269].

\bibitem{Copeland:2003bj}
  E.~J.~Copeland, R.~C.~Myers and J.~Polchinski,
  JHEP {\bf 0406}, 013 (2004)
  [arXiv:hep-th/0312067].

\bibitem{Becker:2005pv}
  K.~Becker, M.~Becker and A.~Krause,
  Phys.\ Rev.\  D {\bf 74}, 045023 (2006)
  [arXiv:hep-th/0510066].

\bibitem{Jackson:2004zg}
  M.~G.~Jackson, N.~T.~Jones and J.~Polchinski,
  JHEP {\bf 0510}, 013 (2005)
  [arXiv:hep-th/0405229].

\bibitem{Jackson:2006rb}
  M.~G.~Jackson,
  JHEP {\bf 0609}, 071 (2006)
  [arXiv:hep-th/0608152].

\bibitem{Jackson:2007hn}
  M.~G.~Jackson,
  arXiv:0706.1264 [hep-th].

\bibitem{Pogosian:2003mz}
  L.~Pogosian, S.~H.~H.~Tye, I.~Wasserman and M.~Wyman,
  Phys.\ Rev.\  D {\bf 68}, 023506 (2003)
  [Erratum-ibid.\  D {\bf 73}, 089904 (2006)]
  [arXiv:hep-th/0304188].

\bibitem{Shlaer:2005ry}
  B.~Shlaer and M.~Wyman,
  Phys.\ Rev.\  D {\bf 72}, 123504 (2005)
  [arXiv:hep-th/0509177].

\bibitem{Leblond:2007tf}
  L.~Leblond and M.~Wyman,
  arXiv:astro-ph/0701427.

\bibitem{Liddle:1993fq}
  A.~R.~Liddle and D.~H.~Lyth,
  Phys.\ Rept.\  {\bf 231}, 1 (1993)
  [arXiv:astro-ph/9303019].

\bibitem{Liddle:1992wi}
  A.~R.~Liddle and D.~H.~Lyth,
  Phys.\ Lett.\  B {\bf 291}, 391 (1992)
  [arXiv:astro-ph/9208007].

\bibitem{Lyth:1996im}
  D.~H.~Lyth,
  Phys.\ Rev.\ Lett.\  {\bf 78}, 1861 (1997)
  [arXiv:hep-ph/9606387].

\bibitem{Baumann:2006cd}
  D.~Baumann and L.~McAllister,
  Phys.\ Rev.\  D {\bf 75}, 123508 (2007)
  [arXiv:hep-th/0610285].

\bibitem{Bean:2007hc}
  R.~Bean, S.~E.~Shandera, S.~H.~Henry Tye and J.~Xu,
  JCAP {\bf 0705}, 004 (2007)
  [arXiv:hep-th/0702107].

\bibitem{Kallosh:2007wm}
  R.~Kallosh and A.~Linde,
  JCAP {\bf 0704}, 017 (2007)
  [arXiv:0704.0647 [hep-th]].

\bibitem{Lidsey:2007gq}
  J.~E.~Lidsey and I.~Huston,
  JCAP {\bf 0707}, 002 (2007)
  [arXiv:0705.0240 [hep-th]].

\bibitem{Peiris:2007gz}
  H.~V.~Peiris, D.~Baumann, B.~Friedman and A.~Cooray,
  arXiv:0706.1240 [astro-ph].

\bibitem{Taylor:2004hha}
  A.~C.~Taylor {\it et al.},
  arXiv:astro-ph/0407148.

\bibitem{Verde:2005ff}
  L.~Verde, H.~Peiris and R.~Jimenez,
  JCAP {\bf 0601}, 019 (2006)
  [arXiv:astro-ph/0506036].

\bibitem{Becker:2005sg}
  K.~Becker, M.~Becker and A.~Krause,
  Nucl.\ Phys.\  B {\bf 715}, 349 (2005)
  [arXiv:hep-th/0501130].

\bibitem{ToAppear}
  to appear

\bibitem{Stewart:1993bc}
  E.~D.~Stewart and D.~H.~Lyth,
  Phys.\ Lett.\  B {\bf 302}, 171 (1993)
  [arXiv:gr-qc/9302019].

\bibitem{Lidsey:1995np}
  J.~E.~Lidsey, A.~R.~Liddle, E.~W.~Kolb, E.~J.~Copeland, T.~Barreiro and M.~Abney,
  Rev.\ Mod.\ Phys.\  {\bf 69}, 373 (1997)
  [arXiv:astro-ph/9508078].

\bibitem{Peiris:2003ff}
  H.~V.~Peiris {\it et al.}  [WMAP Collaboration],
  Astrophys.\ J.\ Suppl.\  {\bf 148}, 213 (2003)
  [arXiv:astro-ph/0302225].

\bibitem{Kinney:2006qm}
  W.~H.~Kinney, E.~W.~Kolb, A.~Melchiorri and A.~Riotto,
  Phys.\ Rev.\  D {\bf 74}, 023502 (2006)
  [arXiv:astro-ph/0605338].

\bibitem{Chongchitnan:2006pe}
  S.~Chongchitnan and G.~Efstathiou,
  Phys.\ Rev.\  D {\bf 73}, 083511 (2006)
  [arXiv:astro-ph/0602594].

\bibitem{Efstathiou:2006ak}
  G.~Efstathiou and S.~Chongchitnan,
  Prog.\ Theor.\ Phys.\ Suppl.\  {\bf 163}, 204 (2006)
  [arXiv:astro-ph/0603118].

\bibitem{Friedman:2006zt}
  B.~C.~Friedman, A.~Cooray and A.~Melchiorri,
  Phys.\ Rev.\  D {\bf 74}, 123509 (2006)
  [arXiv:astro-ph/0610220].

\bibitem{Xia:2007gz}
  J.~Q.~Xia, H.~Li, G.~B.~Zhao and X.~Zhang,
  arXiv:0708.1111 [astro-ph].

\bibitem{Witten:1996mz}
  E.~Witten,
  Nucl.\ Phys.\ B {\bf 471}, 135 (1996)
  [arXiv:hep-th/9602070].

\bibitem{Curio:2000dw}
  G.~Curio and A.~Krause,
  Nucl.\ Phys.\ B {\bf 602}, 172 (2001)
  [arXiv:hep-th/0012152]. \\
  A.~Krause,
  Fortsch.\ Phys.\  {\bf 49}, 163 (2001). \\
  G.~Curio and A.~Krause,
  Nucl.\ Phys.\ B {\bf 693}, 195 (2004)
  [arXiv:hep-th/0308202].

\bibitem{Moore:2000fs}
  G.~W.~Moore, G.~Peradze and N.~Saulina,
  Nucl.\ Phys.\ B {\bf 607}, 117 (2001)
  [arXiv:hep-th/0012104].

\bibitem{Lima:2001jc}
  E.~Lima, B.~A.~Ovrut, J.~Park and R.~Reinbacher,
  Nucl.\ Phys.\ B {\bf 614}, 117 (2001)
  [arXiv:hep-th/0101049].

\bibitem{Curio:2001qi}
  G.~Curio and A.~Krause,
  Nucl.\ Phys.\ B {\bf 643}, 131 (2002)
  [arXiv:hep-th/0108220].

\bibitem{Curio:2006dc}
  G.~Curio and A.~Krause,
  Phys.\ Rev.\  D {\bf 75}, 126003 (2007)
  [arXiv:hep-th/0606243].

\bibitem{Curio:2005ew}
  G.~Curio, A.~Krause and D.~L\"ust,
  Fortsch.\ Phys.\  {\bf 54}, 225 (2006)
  [arXiv:hep-th/0502168].

\bibitem{Krause:2007gj}
  A.~Krause,
  Phys.\ Rev.\ Lett.\  {\bf 98}, 241601 (2007)
  [arXiv:hep-th/0701009].

\bibitem{Becker:2004gw}
  M.~Becker, G.~Curio and A.~Krause,
  Nucl.\ Phys.\  B {\bf 693}, 223 (2004)
  [arXiv:hep-th/0403027].

\bibitem{Buchbinder:2003pi}
  E.~I.~Buchbinder and B.~A.~Ovrut,
  Phys.\ Rev.\  D {\bf 69}, 086010 (2004)
  [arXiv:hep-th/0310112].

\bibitem{Buchbinder:2004im}
  E.~I.~Buchbinder,
  Phys.\ Rev.\  D {\bf 70}, 066008 (2004)
  [arXiv:hep-th/0406101].

\bibitem{Anguelova:2005jr}
  L.~Anguelova and D.~Vaman,
  Nucl.\ Phys.\  B {\bf 733}, 132 (2006)
  [arXiv:hep-th/0506191].

\bibitem{Anguelova:2006qf}
  L.~Anguelova and K.~Zoubos,
  Phys.\ Rev.\  D {\bf 74}, 026005 (2006)
  [arXiv:hep-th/0602039].

\bibitem{Braun:2006th}
  V.~Braun and B.~A.~Ovrut,
  JHEP {\bf 0607}, 035 (2006)
  [arXiv:hep-th/0603088].

\bibitem{Correia:2006vf}
  F.~P.~Correia, M.~G.~Schmidt and Z.~Tavartkiladze,
  Nucl.\ Phys.\  B {\bf 763}, 247 (2007)
  [arXiv:hep-th/0608058].

\bibitem{Gray:2007mg}
  J.~Gray, A.~Lukas and B.~Ovrut,
  arXiv:hep-th/0701025.

\bibitem{Correia:2007sv}
  F.~P.~Correia and M.~G.~Schmidt,
  arXiv:0708.3805 [hep-th].

\bibitem{Lucchin:1984yf}
  F.~Lucchin and S.~Matarrese,
  Phys.\ Rev.\  D {\bf 32}, 1316 (1985).

\bibitem{Liddle:1998jc}
  A.~R.~Liddle, A.~Mazumdar and F.~E.~Schunck,
  Phys.\ Rev.\  D {\bf 58}, 061301 (1998)
  [arXiv:astro-ph/9804177].

\bibitem{Ashoorioon:2006wc}
  A.~Ashoorioon and A.~Krause,
  arXiv:hep-th/0607001.

\bibitem{Kosowsky:1995aa}
  A.~Kosowsky and M.~S.~Turner,
  Phys.\ Rev.\  D {\bf 52}, 1739 (1995)
  [arXiv:astro-ph/9504071].

\bibitem{Alabidi:2006fu}
  L.~Alabidi,
  JCAP {\bf 0702}, 012 (2007)
  [arXiv:astro-ph/0608287].

\bibitem{Spergel:2006hy}
  D.~N.~Spergel {\it et al.}  [WMAP Collaboration],
  arXiv:astro-ph/0603449.

\bibitem{Smith:2005mm}
  T.~L.~Smith, M.~Kamionkowski and A.~Cooray,
  Phys.\ Rev.\  D {\bf 73}, 023504 (2006)
  [arXiv:astro-ph/0506422].

\bibitem{Boyle:2007zx}
  L.~A.~Boyle and A.~Buonanno,
  arXiv:0708.2279 [astro-ph].

\bibitem{Lyth:1984yz}
  D.~H.~Lyth,
  Phys.\ Lett.\  B {\bf 147}, 403 (1984)
  [Erratum-ibid.\  B {\bf 150}, 465 (1985)].

\bibitem{Olsson:2007he}
  M.~E.~Olsson,
  JCAP {\bf 0704}, 019 (2007)
  [arXiv:hep-th/0702109].

\bibitem{Dimopoulos:2005ac}
  S.~Dimopoulos, S.~Kachru, J.~McGreevy and J.~G.~Wacker,
  arXiv:hep-th/0507205.

\bibitem{Boubekeur:2005zm}
  L.~Boubekeur and D.~H.~Lyth,
  JCAP {\bf 0507}, 010 (2005)
  [arXiv:hep-ph/0502047].

\bibitem{Easther:2006qu}
  R.~Easther, W.~H.~Kinney and B.~A.~Powell,
  JCAP {\bf 0608}, 004 (2006)
  [arXiv:astro-ph/0601276].

\bibitem{Linde:2004kg}
  A.~Linde,
  Phys.\ Scripta {\bf T117}, 40 (2005)
  [arXiv:hep-th/0402051].

\bibitem{Kinney:2007np}
  W.~H.~Kinney,
  AIP Conf.\ Proc.\  {\bf 928}, 3 (2007)
  [arXiv:0706.3699 [astro-ph]].

\bibitem{Singh}
  K.~L.~Panigrahi and H.~Singh,
  arXiv:0708.1679 [hep-th].\\
  H.~Singh,
  arXiv:hep-th/0608032.

\bibitem{Huang:2007qz}
  Q.~G.~Huang,
  arXiv:0706.2215 [hep-th].

\end{thebibliography}
\end{document}